\documentclass{aastex61}
\def\figref#1{Figure~\ref{#1}} 
\def\tabref#1{Table~\ref{#1}} 
\def\eqref#1{Eq.~(\ref{#1})} 
\def\secref#1{Section~\ref{#1}} 

\def\kms{km~${\rm s}^{-1}$}

\def\path_nustar{/Users/naomi/Dropbox/Documents/Tex_documents_mine/apjl_rxj1713_TexPad/make_FigTab/}

\def\rxj{RX J1713.7$-$3946}

\def\slbu{4U 1700$-$377}

\def\casA{Cassiopeia A}
\def\tycho{Tycho's SNR}
\def\g19{G1.9$+$0.3}

\def\chandra{{\it Chandra}}
\def\nustar{{\it NuSTAR}}
\def\xmm{{\it XMM-Newton}}
\def\suzaku{{\it Suzaku}}

\def\uG{\hbox{$\mu{\rm G}$}}

\received{MM DD, YYYY}    \revised{MM DD, YYYY}    \accepted{MM DD, YYYY}
\submitjournal{ApJ}

\shorttitle{\nustar\ observations of SNR \rxj}
\shortauthors{Tsuji et al.}

\begin{document}

\title{\nustar\ Observations of The Supernova Remnant \rxj}

\correspondingauthor{Naomi Tsuji and Yasunobu Uchiyama}
\email{n.tsuji@rikkyo.ac.jp, y.uchiyama@rikkyo.ac.jp}

\author[0000-0002-0786-7307]{Naomi Tsuji}
\affil{Department of Physics, Rikkyo University, 3-34-1 Nishi Ikebukuro, Toshima-ku, Tokyo 171-8501, Japan}

\author{Yasunobu Uchiyama}
\affiliation{Department of Physics, Rikkyo University, 3-34-1 Nishi Ikebukuro, Toshima-ku, Tokyo 171-8501, Japan}

\author{Felix Aharonian}
\affiliation{Dublin Institute for Advanced Studies, 31 Fitzwilliam Place, Dublin 2, Ireland}
\affiliation{Max-Planck-Institut f\"ur Kernphysik, P.O. Box 103980, D 69029 Heidelberg, Germany}

\author{David Berge}
\affiliation{DESY, Platanenallee 6, 15738 Zeuthen, Germany}

\author{Ryota Higurashi}
\affiliation{Department of Physics, Rikkyo University, 3-34-1 Nishi Ikebukuro, Toshima-ku, Tokyo 171-8501, Japan}

\author{Roman Krivonos}
\affiliation{Space Research Institute of the Russian Academy of Sciences (IKI), Moscow, Russia, 117997}

\author{Takaaki Tanaka}
\affiliation{Department of Physics, Kyoto University, Kitashirakawa Oiwake-cho, Sakyo, Kyoto 606-8502, Japan}





\begin{abstract}
The shock waves of supernova remnants (SNRs) are prominent candidates for the acceleration of the Galactic cosmic rays.
SNR \rxj\ is one well-studied particle accelerator in our Galaxy because of its strong non-thermal X-ray and gamma-ray radiation.
We have performed \nustar\ (3--79 keV) observations of the northwest rim of \rxj, where is the brightest part in X-ray and the shock speed is about 4000 \kms.
The spatially resolved X-ray emission from \rxj\ is detected up to 20 keV for the first time.
The hard X-ray image in 10--20 keV is broadly similar to the soft-band image in 3--10 keV.
The typical spectrum is described by power-law model with exponential cutoff with the photon index $\Gamma$=2.15 and the cutoff energy $\varepsilon_c$=18.8 keV.
Using a synchrotron radiation model from accelerated electrons in the loss-limited case, 
the cutoff energy parameter ranges 0.6--1.9 keV, varying from region to region.
Combined with the previous measurement of the shock speed,
the acceleration of electrons is close to the Bohm-limit regime in the outer edge,
while the standard picture of accelerated particles limited by synchrotron radiation in SNR shock is not applicable in the inner edge and the filamentary structure.

\end{abstract}




\section{Introduction} \label{sec:intro}

Shell-type supernova remnant (SNR) \rxj\ is well known for its strong non-thermal X-ray and gamma-ray emission, 
making it to be one of the best-studied particle acceleration sites \citep{Tanaka2008, Abdo2011, HESS2018}.
The association between \rxj\ and SN 393, one of the historical SNRs, has been discussed \citep{Wang1997, Fesen2012}.
Recent measurements of the proper motions in the northwest (NW) and the southeast part of this SNR \citep{TU16,Acero2017}
revealed that the forward shock speed is roughly 4000 \kms.
This suggests that \rxj\ is indeed the remnant of SN 393 and kinematically young, 
implying it is still in the ejecta-dominated phase.
Both the fast shock velocity and the early evolutional phase are consistent with the efficient acceleration of particles in the SNR.

The particle acceleration in \rxj\ is likely in the most efficient regime.
\cite{ZA07} developed an analytical expression for the spectral energy distribution of electron around the SNR shock and the corresponding synchrotron radiation 
in the framework that the energy loss of the accelerated electrons is dominated by the synchrotron cooling (hereafter ZA07 model).
\cite{Tanaka2008} reported that the broadband X-ray spectrum of the entire remnant of \rxj\ with \suzaku\ (XIS and HXD PIN; 0.4--40 keV)
is nicely described by ZA07 model with the cutoff energy parameter ($\varepsilon_0$) of 0.67 keV.
Combined with the shock speed, they reported the acceleration of electrons is in the regime close to the Bohm limit, where the so-called gyro factor is close to unity.

\nustar, {\it Nuclear Spectroscopic Telescope Array} \citep{Harrison2013}, enables us to perform the first time observation of the spatially resolved hard X-ray above 10 keV.
The non-thermal (synchrotron) X-ray has been detected with \nustar\ from some young Galactic SNRs: 
e.g., \casA, \tycho, \g19, and SN1006 \citep{Grefenstette2015,Sato2018, Lopez2015, Zoglauer2015,Aharonian2017,Li2018}.
The spatially-resolved spectral analysis was performed for these SNRs.
In \casA, \tycho, and SN1006, the spectral properties are significantly different from region to region, 
suggesting that the efficiency of acceleration depends on the site.
Although \g19\ shows uniform spectra across the remnant \citep{Zoglauer2015},
the acceleration of electrons is about one order of magnitude slower than Bohm limit, as reported in \cite{Aharonian2017}.

Our \nustar\ observations of \rxj\ are the third example of the detections of the spatially-resolved non-thermal hard X-ray emissions with \nustar\ 
from the synchrotron-dominated SNRs, following \g19\ and SN1006.
The hard X-ray emission from \rxj\ was detected with {\it INTEGRAL}/IBIS in 16--70 keV \citep{Krivonos2007_INTEGRAL} and \suzaku/HXD PIN in 12--40 keV \citep{Tanaka2008}.
However the angular resolution of these X-ray instruments is limited:
{\it INTEGRAL}/IBIS is a coded-mask telescope and
\suzaku/HXD is a non-imaging, collimated instrument with the detector units of silicon PIN diodes and GSO scintillators.
We investigate the spatial distribution of the hard X-ray on arcmin scale using \nustar, which consists of hard X-ray telescopes and detectors.

In this paper, we present the results of our spatially-resolved hard X-ray observations obtained with \nustar\ from the NW rim of \rxj.
\secref{sec:2} summarizes the \nustar\ observations.
The imaging and the spectral analysis are described in \secref{sec:3}.
We discuss the efficiency of particle acceleration in several arcmin-scale regions in the NW in \secref{sec:4}.
The conclusions are presented in \secref{sec:5}.

\section{Observation} \label{sec:2}

With the \nustar\ satellite \citep{Harrison2013}, we have performed observations of the NW region of \rxj\ twice, in 2015 September (hereafter P1) and in 2016 March (P2), 
with exposure times of 50 ks and 57 ks, respectively (\tabref{tab:nustar_dataset}). 
 These data are taken with two focal plane modules, named FPMA and FPMB, on board the \nustar\ satellite, co-aligned with the corresponding optics modules. The energy range of both detecters is 3--79 keV. 

All data are calibrated and screened by using {\tt nupipeline} of \nustar\ {\it Data Analysis Software} (NuSTARDAS version 1.4.1 with CALDB version 20180814), included in HEAsoft version 6.19. 
For the data screening, we use the strictest mode (SAAMODE=STRICT and TENTACLE=YES cut).
This process reduces the effective observation time to 43 ks and 49 ks for P1 and P2, respectively (\tabref{tab:nustar_dataset}).

\begin{deluxetable*}{cccccccc}[ht!]
\tablecaption{Log of \nustar\ observations}
\tablehead{
\colhead{} & \colhead{ObsID} & \colhead{Start date} & \twocolhead{Pointing position}  & \colhead{Positional angle} & \colhead{Exposure} & \colhead{Effective time\tablenotemark{$\dagger$}} \\
\colhead{} & \colhead{} & \colhead{(yyyy-mm-dd)} & \colhead{($\alpha, \delta$)\tablenotemark{$\ddagger$} } &  \colhead{($l, b$)\tablenotemark{$\ast$} } &  \colhead{(degree)} & \colhead{(ks)} & \colhead{(ks)}
}
\startdata
\hline
P1 & 40111001002  &    2015-09-27 &  257.86, $-$39.52   &  347.29, $+$0.001 & 343.3  &  50 & 43   \\   
P2 & 40111002002  &    2016-03-30   &  257.93, $-$39.58   &  347.28, $-$0.077 & 165.6  & 57 & 49   \\
\hline
T CrB & 30101046002  &    2015-09-23 &  239.85, $+$25.90   &   42.34, $+$48.18  &319.5  &  80 & 63   \\   
\enddata
\tablenotetext{\dagger}{After screening with SAAMODE=STRICT and TENTACLE=YES cut}
\tablenotetext{\ddagger}{RA and Dec in unit of degree}
\tablenotetext{\ast}{Galactic longitude and latitude in unit of degree}
\label{tab:nustar_dataset}
\end{deluxetable*}

\section{Analysis and Results} \label{sec:3}

\subsection{Detailed analysis of \nustar}
\label{sec:analysis}
\nustar\ has side aperture contamination.
Due to the separation between the telescope modules and the detectors and 
its openness to the sky window, 
the X-ray emissions from the outside of the field of view (FoV) can be detected directly in the FPMs without passing through the optic modules.
They are referred as stray lights. 
They come from the region within radius of 2.5\degr\ from the center of the FoV.
The stray lights are caused by bright X-ray sources located in the vicinity of the target point,
Cosmic X-ray Background (CXB), 
and Galactic Ridge X-ray Emission (GRXE; \cite{Revnivtsev2006,Krivonos2007_GRXE,Yuasa2012}) when pointing the Galactic plane.
This causes complex distributions of the background.

\begin{figure*}[h!] 
\plotone{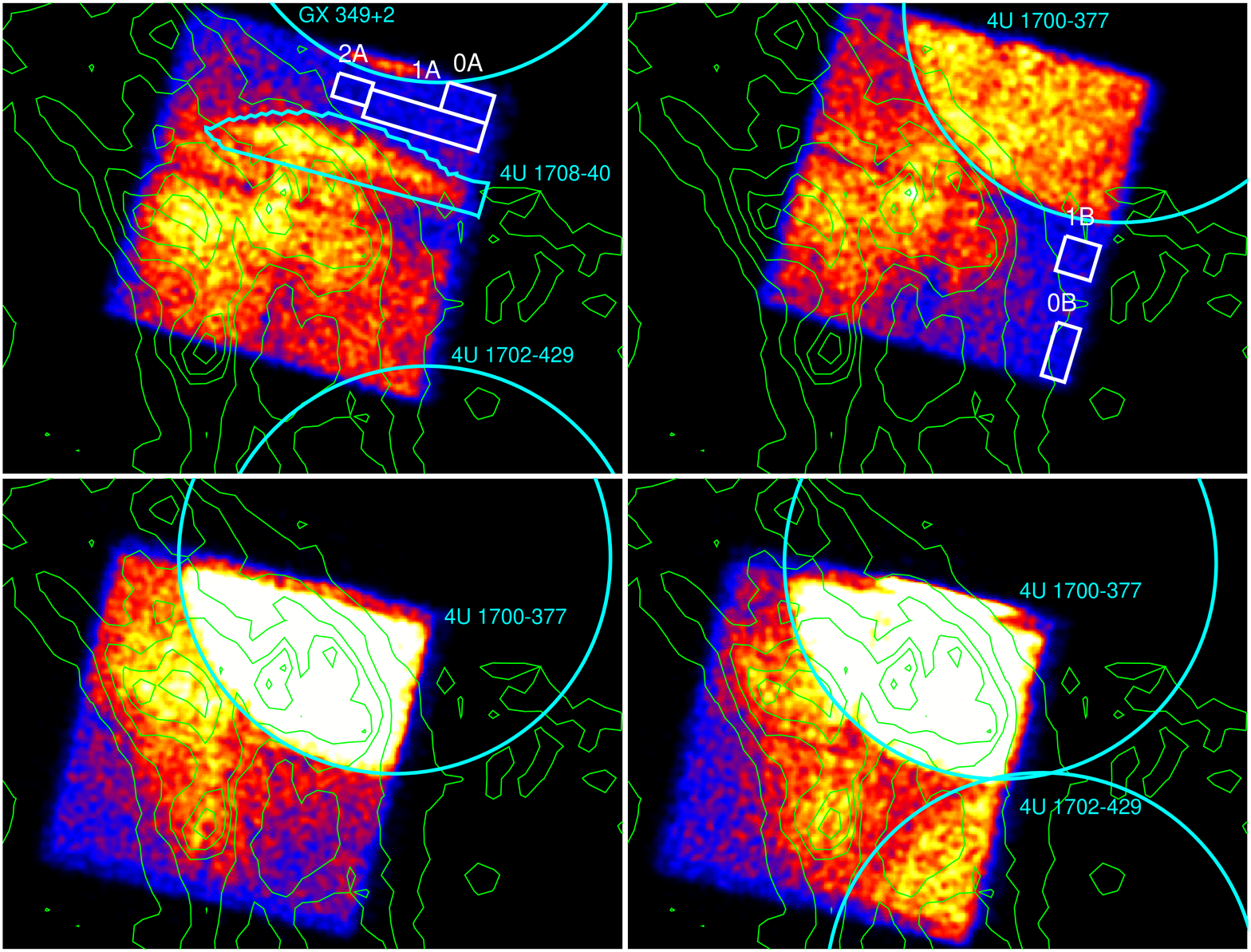}
\plottwo{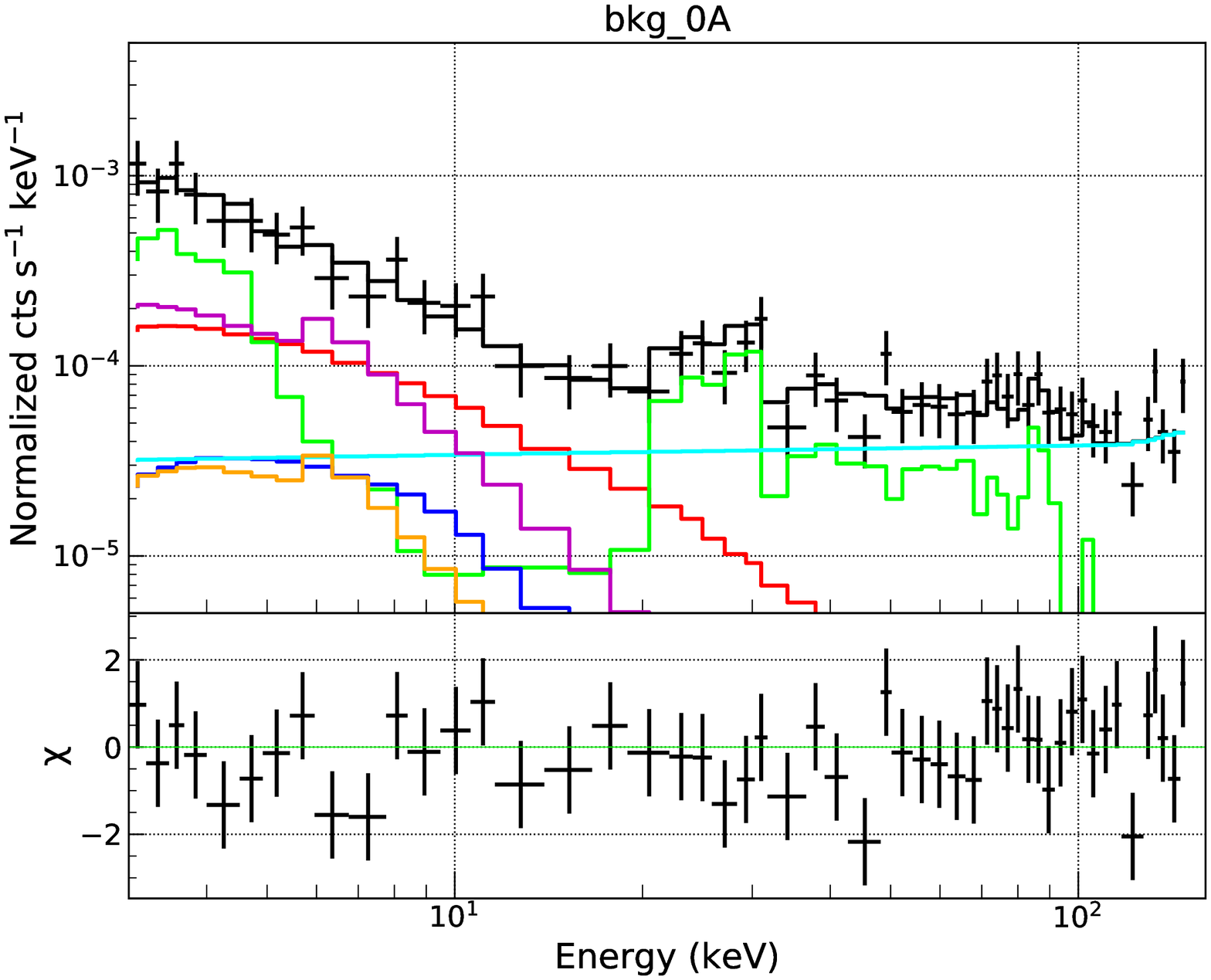}{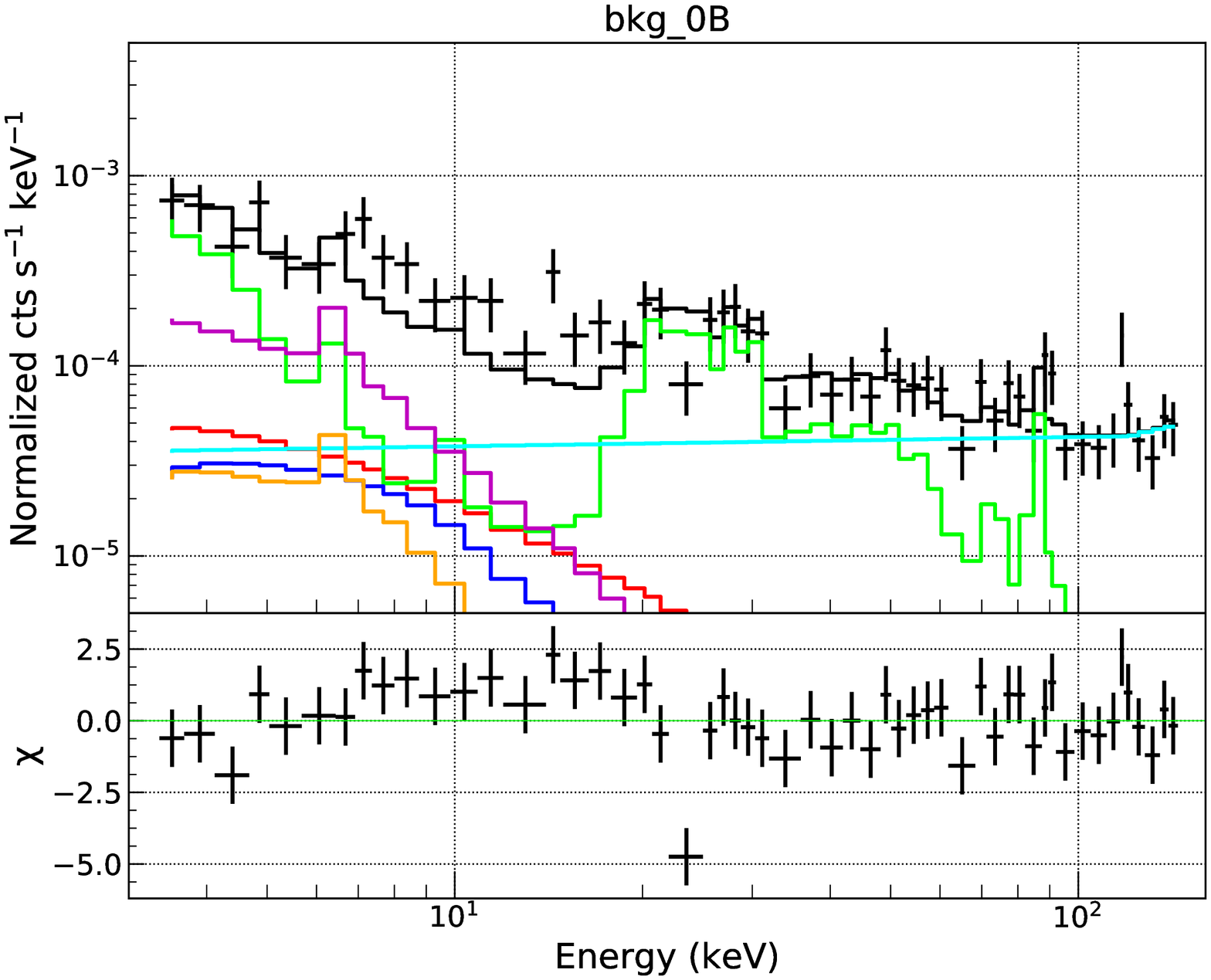}
\caption{
Top left: The count map of P1-FPMA. 
The cyan regions illustrate the expected stray-light patterns from known X-ray sources in the vicinity of \rxj.
The background regions are shown with white boxes.
The contours are taken from the \xmm\ image in 0.5--8 keV.
Top right: Same as the top left for P1-FPMB. 
Middle left: Same as the top left for P2-FPMA. 
Middle right: Same as the top left for P2-FPMB. 
Bottom left: The background spectrum of 0A region with nuskybgd model plus GRXE. 
The black, red, blue, green, cyan lines respectively indicate the total background, aCXB, fCXB, Inst, and Intn, included in nuskybgd.
The magenta and orange lines are aGRXE and fGRXE, respectively.
Bottom right: Same as bottom left for the spectrum of 0B.
}
\label{fig:nuskybgd_fig1}
\end{figure*}

Since \rxj\ is located on the Galactic plane, there are some bright X-ray sources as possible stray-light contaminators. 
For example, the stray light from HMXB \slbu\ dominates in all X-ray energy bands except for P1 observation with FPMA.
In addition, there exist stray-light contaminations from GX 349$+$2, 4U 1708$-$40 in P1-FPMA, 
and 4U 1702$-$429 in P1-FPMA and P2-FPMB.
\figref{fig:nuskybgd_fig1} illustrates the regions contaminated by the stray lights from these X-ray sources.
We exclude these contaminated regions for the following imaging and spectral analysis.

Another component to cause stray-light contaminations is CXB. 
In addition to the focused CXB emission passing through the optics,
CXB coming from the outside of the FoV without passing through the telescopes are detected non-uniformly in the FPMs.
It should be noted that we need to take these nonuniform distributions into account
because \rxj\ is extended across most of the FoV.
In order to deal with this issue, we use the toolkit named ``nuskybgd"\footnote{https://github.com/NuSTAR/nuskybgd} \citep{Wik2014}.

Thanks to nuskybgd, we can construct the background for any region in which we are interested. 
The general usage of nuskybgd is as follows.
First we extract the spectra of ``source-free" regions outside the SNR, where any stray lights from X-ray sources should not be contained. 
In nuskybgd, the model consists of the following four components: 
1.) ``aCXB" to describe the stray-light CXB through the aperture, 
2.) ``fCXB" to describe the focused CXB, 
3.) ``Inst" to describe instrument line emissions and reflected solar X-rays, and 
4.) ``Intn" to describe instrument Compton scattered continuum emissions.
Using {\tt nuskybgd\_fitab} included in nuskybgd, we can fit the background spectrum by the model above.
Based on the best-fit parameters, 
the background image and the background spectrum for an arbitrary region in the FoV are respectively produced by using {\tt nuskybgd\_image} and {\tt nuskybgd\_spec} in nuskybgd.

GRXE is not negligible because \rxj\ is located on the Galactic plane.
In this paper, we use ``apec" model in XSPEC with the electron temperature ($kT$) of 8 keV for GRXE.
Indeed, off-set observations near \rxj\ with {\it Suzaku} showed that 
CXB and GRXE, which is described by apec with $kT$=7.4--8.8 keV, are dominant above 3 keV \citep{Katsuda2015}.
Similar to CXB, GRXE has also a focused component (``fGRXE") and an unfocused component (``aGRXE").
We assume fGRXE is uniform in the vicinity of \rxj, and fix the normalization to the value derived from {\it Suzaku} observations in \cite{Katsuda2015}.
The distribution of aGRXE, however, is not well understood yet.
Unlike aCXB coming from the flat sky, 
modeling aGRXE distribution on the detectors is a difficult task, 
since it mainly comes from the nonuniform Galactic bulge.
Here we assume aGRXE is simply uniform on the focal plane modules, 
and consider systematic uncertainties associated with this approximation in \secref{sec:spectrum}.

Our background estimation of \rxj\ with \nustar\ is as follows.
We extract the background spectra from the regions referred to 0A--2A in FPMA and 0B--1B in FPMB, shown in \figref{fig:nuskybgd_fig1}.
To estimate the background of FPMA, the spectra of 0A--2A are jointly fitted, tying the normalizations between them based on the relative brightnesses expected from the known nonuniform distribution. 
The spectra of 0B--1B are also jointly fitted to estimate the background of FPMB. 
Dividing the background region into a few pieces, such as 0A, 1A, and 2A, helps us to better estimate the background, as recommended in \cite{Wik2014}.
The background model consists of aCXB, fCXB, Inst, Intn, aGRXE and fGRXE.
We use nuskybgd for CXB and the instrumental components,
while we add GRXE manually since the modelling of GRXE is not yet developed in nuskybgd.
When fitting the background spectra, the normalizations of fCXB and aCXB are fixed to the following values.
The normalization of fCXB is fixed to the value defined in the literature \citep{Boldt1987}.
Since CXB is uniform over the sky, the normalization of aCXB is derived from a high-latitude \nustar\ observation, where no GRXE is contained.
The observation of a symbiotic star T CrB with the latitude of 48.18\degr\ is selected for this purpose (\tabref{tab:nustar_dataset}).
We fit the background spectra of T CrB, which are extracted from the same detector region as that of \rxj, using nuskybgd in order to get the level of aCXB.
The obtained normalization is then used for the normalization of aCXB for the observations of \rxj.

The spectra of 0A and 0B are shown with all the components of the background model in \figref{fig:nuskybgd_fig1}.
The background spectra are well-fitted by our background model.
It is notable that the background spectrum of 0B is extracted from the region where aCXB is weakest in FPMB, making the component of aCXB lower.
The free parameters in low energy band and high energy band are, respectively, the normalization of aGRXE and the instrument emissions.
The normalization of aGRXE is roughly obtained to be $\int n_\mathrm{e} n_\mathrm{H} d\ell = 7\times 10^{19}~ \mathrm{cm}^{-5}$ for both FPMA and FPMB,
based on the assumption of the distance of 1 kpc \citep{Fukui2003}.

In order to construct the background spectrum for the region of interest,  
we use {\tt nuskybgd\_spec} in nuskybgd for aCXB, fCXB, Inst and Intn components. 
It takes into account the nonuniform distribution of the background, i.e. the background normalization for the region of interest is weighted by the nonuniform distribution on the detectors.
On the other hand, for fGRXE and aGRXE we simply rescale the normalization by the size of the source region,  
assuming the flat GRXE distribution across the FoV.
We discuss the validity of this assumption in \secref{sec:spectrum}.

For P2 epoch, the background is estimated based on the best-fit parameters of the background of P1 observation.
Since the emission of \rxj\ is extended across the entire FoV of P2,
there is no region to extract for the background estimation.
We assume that the background with \nustar\ is approximately stable over the period of these two observations, P1 (2015 September) and P2 (2016 March).
The background of an arbitrary region in P2 observation is generated in the same procedure as P1, as mentioned above.

\subsection{Image} 
\label{sec:image}
\begin{figure*}[t!]
\plotone{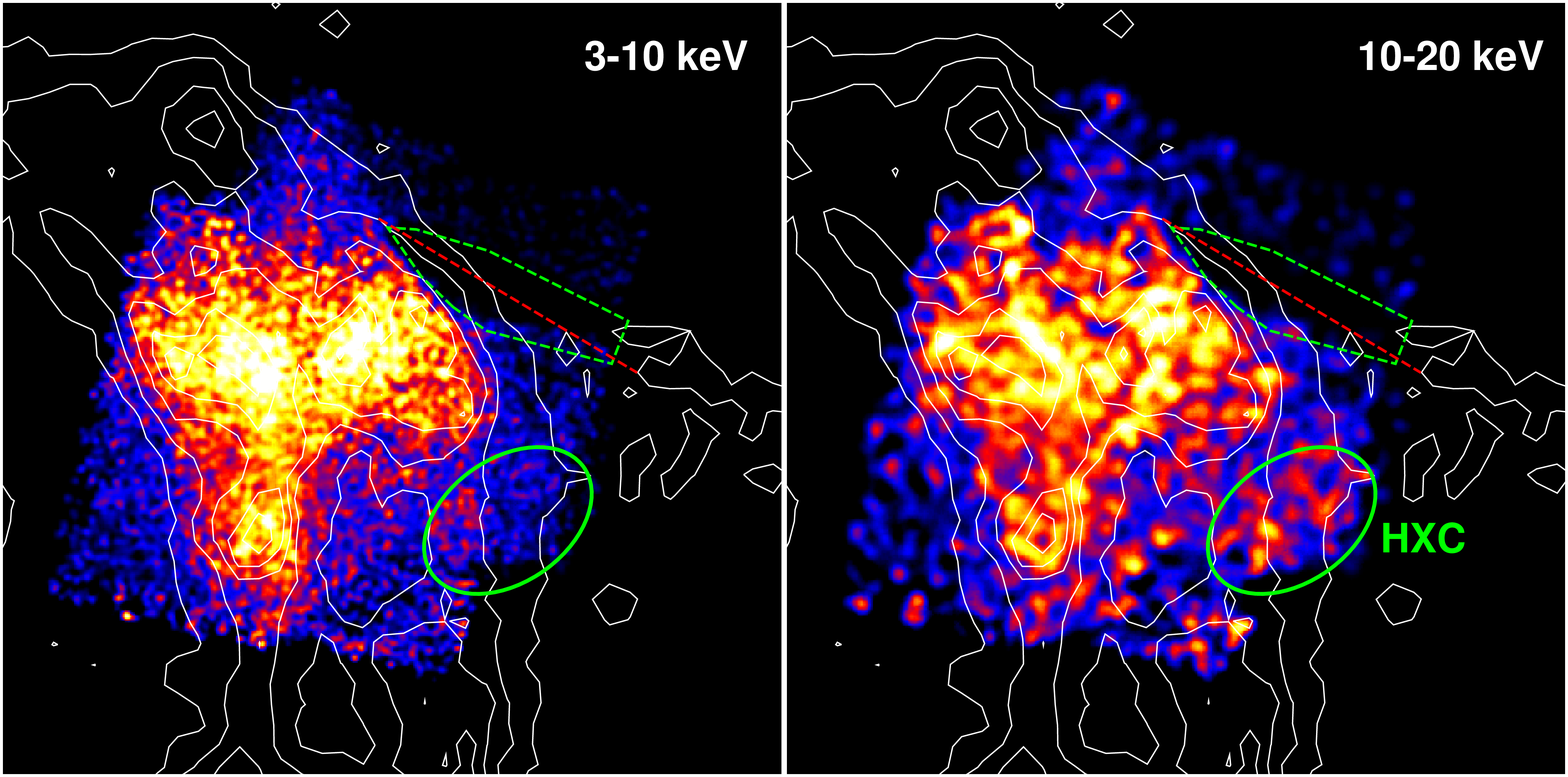}
\caption{
Background-subtracted and exposure-corrected images with \nustar\ (left: 3--10 keV, right: 10--20 keV).
The images of P1 and P2 are added to each other. 
The contours are taken from \xmm\ observations in 0.5--8 keV. 
The images include the contribution of GRXE.
The regions contaminated by the stray lights from X-ray sources, shown in \figref{fig:nuskybgd_fig1}, are removed.
The region with the green dashed line shows the region with no photon due to the removal of the stray lights.
The position of HXC is illustrated with the green ellipse.
}
\label{fig:nustar_image}
\end{figure*}

In \figref{fig:nustar_image}, we present the images in energy bands of 3--10 keV and 10--20 keV.
These are background-subtracted and exposure-corrected images,
where the background images are constructed using nuskybgd 
and the exposure maps are generated by {\tt nuexpomap} in NuSTARDAS with no vignetting correction, 
after removing the regions contaminated by the stray lights from X-ray sources. 
We note that GRXE is included in these images, while the other background, CXB and the instrument background, are subtracted.
The observations of the two epochs and the two FPMs are combined.
The 3--10 keV image is roughly consistent with the soft X-rays in 0.5--8 keV taken by \xmm\ shown with the contour. 

We find an interesting feature, which is referred to HXC (Hard X-ray Component), at the position of the green ellipse that is centered at ($\alpha_{\rm J2000}$, $\delta_{\rm J2000}$) $=$
(17\fh11\fm8\farcs68, $-$39\fd37\fm7\farcs23) in the 10--20 keV image,
while the other part of the hard X-ray image is roughly in agreement with the soft X-rays.
This region is faint in the lower X-ray band, making its spectrum extremely hard ($\Gamma\sim 1$).
It should be noted that the region around HXC is not contaminated by any stray lights from X-ray sources.
It remains as a future work to confirm the presence of HXC and interpret the physical meaning.

\subsection{Spectrum}
\label{sec:spectrum}
\nustar, for the first time, allows us to perform spatially resolved spectroscopy in the hard X-ray band above $\sim$10 keV. 
We investigate the hard X-ray spectral distribution for different regions in the NW of \rxj.
All the spectra are produced using {\tt nuproducts} in NuSTARDAS 
with parameter ``extended" set to ``yes".
All the background spectra are generated by nuskybgd and adding GRXE, as described in \secref{sec:analysis}.
The spectral fitting is performed with XSPEC version 12.9.0.

The energy range of \nustar\ spectra for the spectral analysis is set to be 3--20 keV due to the poor statistics above 20 keV,
caused by stronger instrumental background. 
Combining the \nustar\ spectra with the 0.5--7 keV spectra obtained with \chandra,
we perform \chandra\ $+$ \nustar\ joint fitting.
For \chandra\ data, we use six observations of \rxj\ NW with ObsID of 736, 6370, 10090, 10091, 10092, and 12671. 
The data are processed using CALDB version 4.7.6 in CIAO version 4.9.
All the spectra of different epochs with \chandra\ are combined using {\tt addascaspec} in HEAsoft.

\begin{figure}[t!]
\begin{center}
\plotone{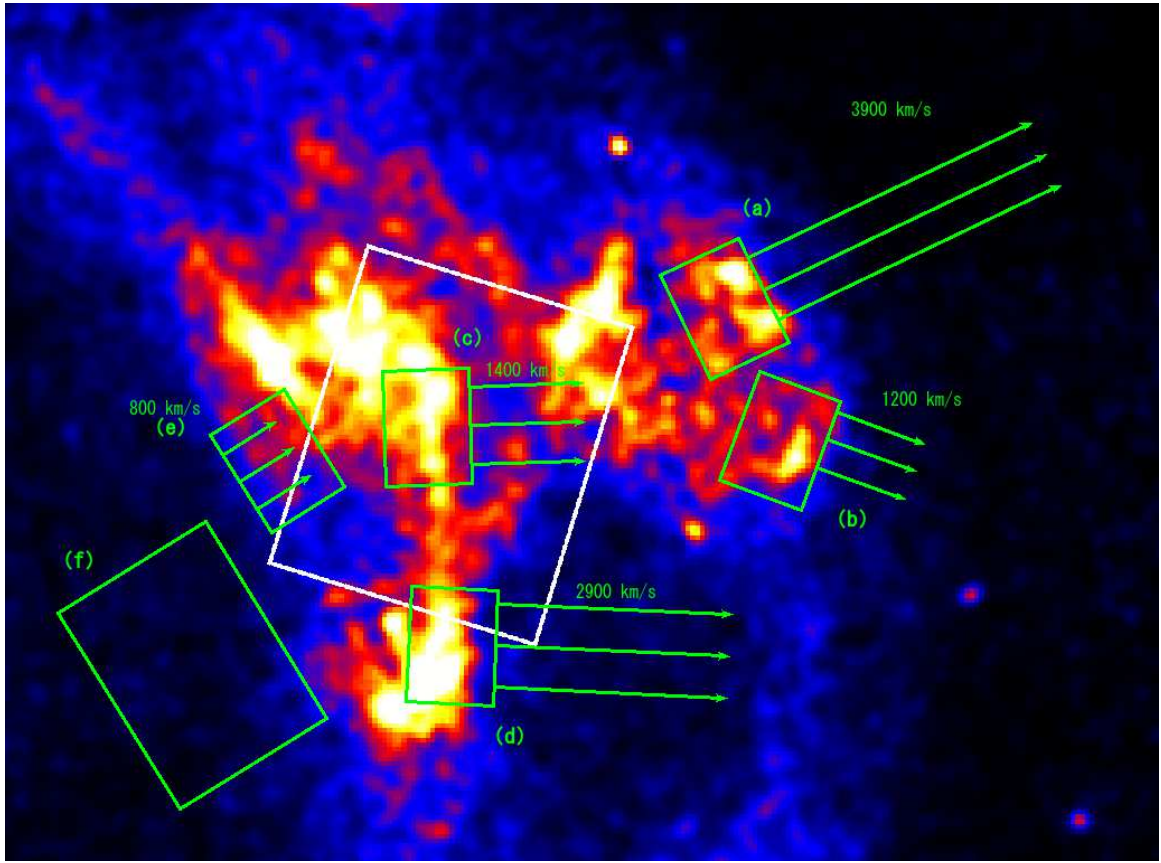}
 \caption{
Locations of the large box and the small box (a) -- (f), shown with the white box and the green boxes, respectively. 
The image is 0.5--8 keV with \xmm. The color is shown in squared scale.
The length of an arrow indicates the shock speed measured by \chandra\ \citep{TU16}.
} 
\label{fig:regions}
\end{center}
\end{figure}

The spectral fitting is performed for the integrated region (large box) and spatially resolved arcmin-scale regions (small box (a)--(e)), as shown in \figref{fig:regions}.
We use two conventional models which are power-law model and power-law with exponential cutoff model,
as well as ZA07 model.
\cite{ZA07} obtained the analytical expressions of the synchrotron radiation spectrum from accelerated electrons  
in the framework that assumes electrons accelerated via DSA are cooled by synchrotron emissions. 
The analytical solution of the electron spectrum and the synchrotron-photon spectrum in the downstream region are respectively,
\begin{eqnarray}
\frac{dN_e}{dE} &\propto & \left(\frac{E}{E_0}\right)^{-3} 
\Bigg[ 
\left\{ 1+0.523\left( \frac{E}{E_0} \right)^{\frac{9}{4}} \right\}^{2}
-  0.0636 \left(\frac{E}{E_0}\right)^{2} \left\{ 1+1.7\left( \frac{E}{E_0} \right)^3 \right\}^{\frac{5}{6}}
\Bigg] 
\exp \Bigg[ -\left(\frac{E}{E_0}\right)^2 \Bigg] \label{eq:ZA07_model_electron} , \\ 
\frac{dN}{d\varepsilon} &\propto & \left(\frac{\varepsilon}{\varepsilon_0}\right)^{-2} \Bigg[ 1+0.38\left( \frac{\varepsilon}{\varepsilon_0} \right)^{\frac{1}{2}} \Bigg]^{\frac{11}{4}}\exp \Bigg[ -\left(\frac{\varepsilon}{\varepsilon_0}\right)^{\frac{1}{2}}\Bigg] , \label{eq:ZA07_model_synch} 
\end{eqnarray}
which are same as Eq. (28) and Eq. (37) in \cite{ZA07}.
Here $E_0$ and $\varepsilon_0$ are the cutoff energy parameters of the electron and the synchrotron X-ray, respectively. 
Note that the cutoff energy ($\varepsilon_c$) in the cutoff power-law model is about one order of magnitude higher than $\varepsilon_0$ in ZA07 model.
This is because $\varepsilon_c$ gives  $e^{-1}$ deviation from the power law, while $\varepsilon_0$ does not due to the additional function to the cutoff power law (\eqref{eq:ZA07_model_synch}).
The interstellar absorption is taken into account for all the models using ``TBabs" model in XSPEC.

We take the uncertainty of the background spectrum into consideration as a systematic error.
When fitting the background spectra with {\tt nuskybgd\_fitab} in nuskybgd, the best-fit parameters have about 10\% error.
Therefore, we check how much the source spectrum from the region of interest  depends on the choice of the background spectrum, by changing the normalization.
If we change the normalization of the total background spectrum by 10\%, the source spectrum changes by 10\%.
This uncertainty is used as the systematic error in the following.
It should be noted that the uncertainty of the aGRXE component could be much larger than the statistical error
because we simply assume the uniform distribution of aGRXE in \secref{sec:analysis}.
If we change only the normalization of the aGRXE spectrum by a factor of 3, the source spectrum changes by 10\%.
Thus our assumption on the uniform aGRXE does not largely affect the source spectrum as long as aGRXE is non-uniformly distributed by a factor of $\leq 3$.
We note that in fact CXB is expected to be fluctuated by $\sim$20\% in the FoV of \nustar\ \citep{Moretti2009}. 
Changing the normalization of aCXB, we also checked this uncertainty does not have a large effect on the source spectrum,
i.e., the cutoff energy changes by $\sim$15\%, and the other parameters remain same within $\sim$2\% for 20\% fluctuation of aCXB.

\begin{figure*}[t!]
\begin{center}
\plottwo{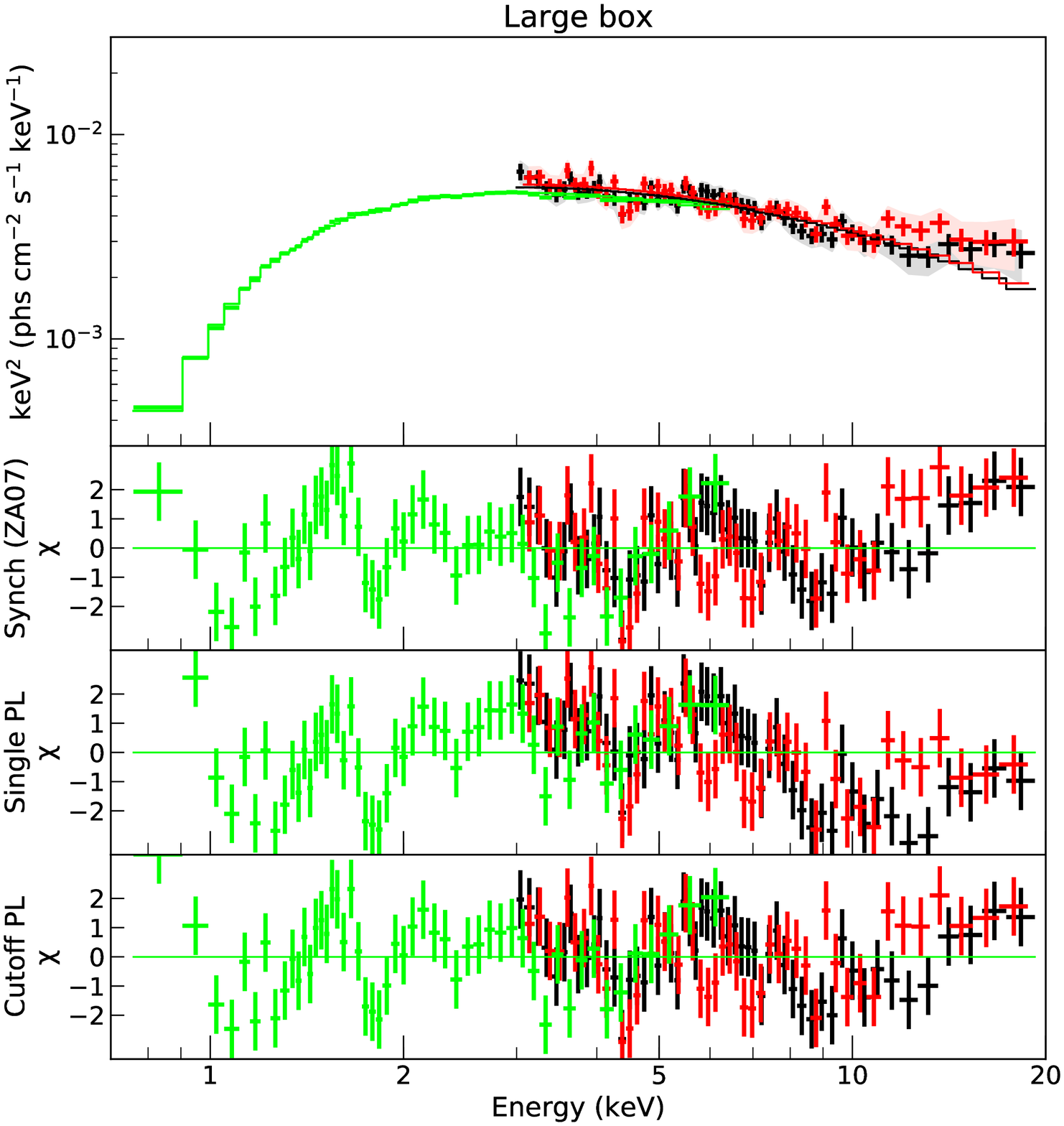}{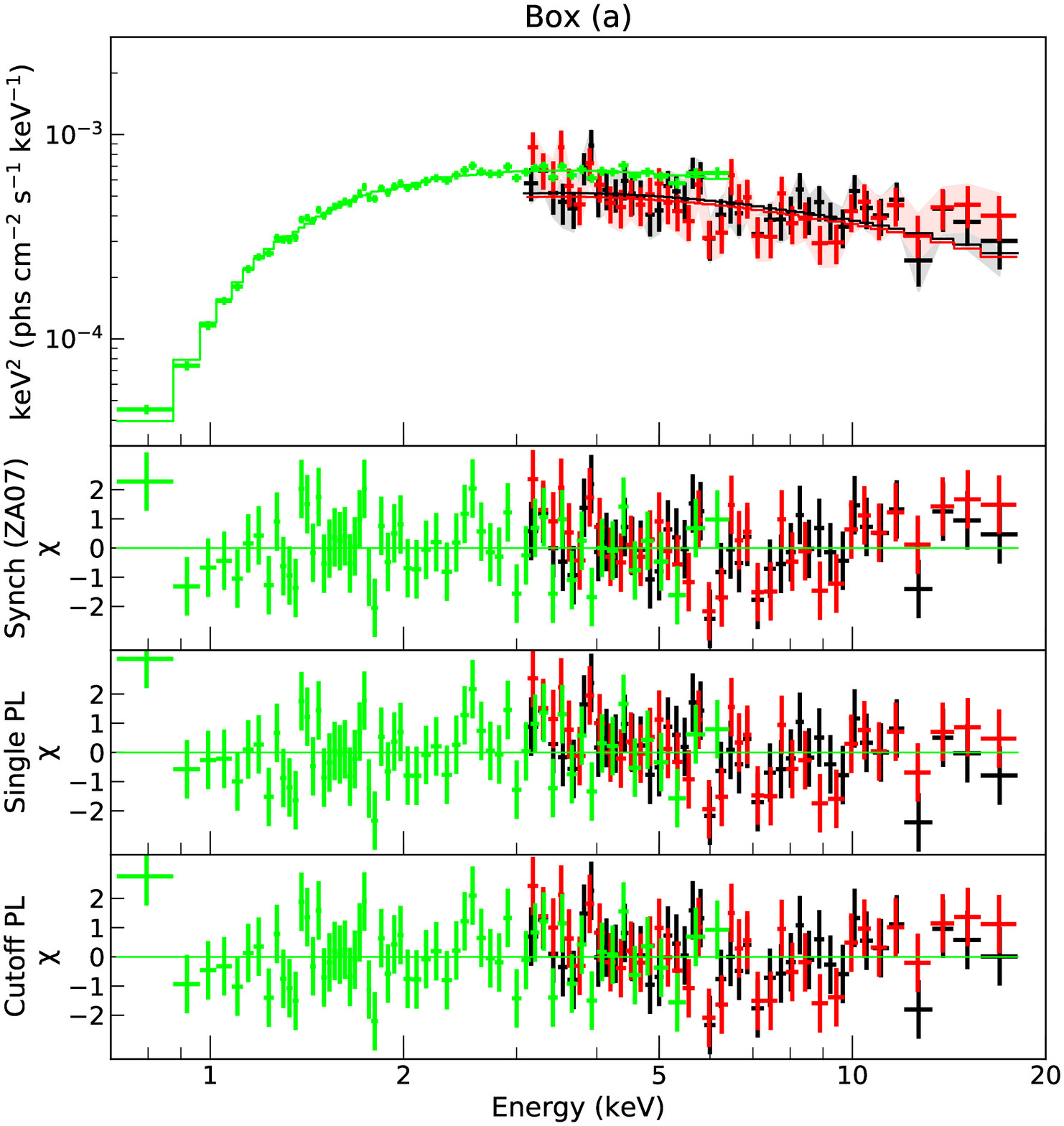}
 \caption{
 The spectrum of large box (left) and box (a) (right).
 Black, red and green plots respectively show the spectrum by FPMA,  FPMB, and \chandra.
The top panel is the spectrum with the best-fit ZA07 model. 
The shadows represent the systematic uncertainties.
The residuals from ZA07 model, single power-law model, and cutoff power-law model are shown in order from the second panel to the bottom, respectively.
 }
\label{fig:nustar_spec}
\end{center}
\end{figure*}

\figref{fig:nustar_spec} (left) shows the spectrum of the large box.
The angular size of the box is 5.7\arcmin$\times$4.8\arcmin, which corresponds to 1.7$\times$1.4 pc$^2$  assuming the distance is 1 kpc \citep{Fukui2003}.
This large box represents the typical spectrum in the NW of \rxj.
The best-fit parameters are listed in \tabref{tab:fit}.
Our analysis excludes the pure power-law model and requires the cutoff in the spectrum.
The cutoff shape is described with the photon index $\Gamma=2.15\pm0.04\pm0.02$ and $\varepsilon_c$= 18.8~$^{+4.2}_{-3.0}~ ^{+2.6}_{-2.0}$ keV for the simple cutoff  power-law model, and $\varepsilon_0$= 1.14~$\pm0.06\pm0.02$ keV for ZA07 model,
where the first and second error indicate the statistic and the systematic error, respectively.
Note that the systematic error is less than the statistical error.
The cutoff power-law model seems to be slightly favoured, as inferred from the chi-squared value.
However we should be cautious about the coupling of the cutoff energy and the photon index.
Both $\varepsilon_c$ and $\Gamma$ are higher than the values reported in \cite{Tanaka2008}.
Furthermore, the best-fit value of the photon index ($\Gamma$=2.15) is larger than 2, 
which is expected for the synchrotron X-ray spectrum radiated from the electron which the acceleration is limited by the synchrotron cooling.
If we fix $\Gamma$ to the theoretically predicted value of 2, the chi-squared value becomes much larger.
ZA07 model gives a better fit than the CPL model with $\Gamma=2$.

We report the first result of arcmin-scale spectral distribution in the hard X-ray band of the NW shell of \rxj.
Six small regions, box (a)--(f), are selected for the following reasons.
Box (a)--(f) include clearly edge-like and filament-like structures. 
We have previously measured the proper motion velocities in these structures \citep{TU16} 
and found that the speeds are significantly different from region to region.
To test the relation between the shock speed and the cutoff energy, 
we define box (a)--(f), which correspond to the regions used in \cite{TU16}.
The size of each box is 1.5\arcmin$\times$2\arcmin\ (0.44$\times$0.58 pc$^2$) except for box (f) with the size of 3\arcmin$\times$4\arcmin\ (0.88$\times$1.16 pc$^2$).
The spectrum of box (a) is shown in \figref{fig:nustar_spec} 
and the best-fit parameters are listed in \tabref{tab:fit}.

\figref{fig:cutoff} demonstrates the relation between the shock speed obtained by \cite{TU16} and 
the cutoff energy parameter derived from the spectral fitting with ZA07 model for each small box. 
The shock velocities are assumed to be same as the proper motion velocities:
3900$\pm$300 \kms\ at box (a),
1200$\pm$300 \kms\ at box (b),
1400$\pm$200 \kms\ at box (c),
2900$\pm$200 \kms\ at box (d), and
800$\pm$300 \kms\ at box (e) and box (f),
assuming the distance is 1 kpc \citep{Fukui2003}.
The measured speeds contain the uncertainties of being projected onto the line of sight.
However these uncertainties can be small because boxes (a)--(f) are located in outer regions of the shell and the radial component is expected to be dominant in these regions.
The projection-corrected speeds, plotted with the open markers in \figref{fig:cutoff}, are inferred from taking into account the projection effects, i.e. the line-of-sight velocity components, 
assuming the spherical shell expansion.

\begin{deluxetable*}{cccccccc}[t!]
\tablecaption{The best-fit parameters of the spectral analysis}
\small
\tablehead{
  \colhead{Box}   &   \colhead{Model}  &   \colhead{$N_H$}    &   \colhead{$\Gamma$}    &   \colhead{$\varepsilon_c$/$\varepsilon_0$}  & \colhead{Flux$_{3-20~ {\rm keV}}$ \tablenotemark{$\ast$} } & \colhead{$\chi^2$} & \colhead{(d.o.f)} \\
\colhead{} & \colhead{}  & \colhead{$\left(10^{22}~{\rm cm}^{-2}\right)$} & \colhead{} & \colhead{(keV)}     & \colhead{(10$^{-12}$ erg/cm$^2$/s) } & \colhead{} &\colhead{}
}
\startdata
 Large\tablenotemark{$\dagger$}  & PL  &   $0.815\pm0.011$ &   $2.32\pm0.02$   &  -- &  13.4$\pm0.1$  & 92.9 & 48 \\ 
 Large\tablenotemark{$\ddagger$}  & PL  &  0.815; fixed   &  2.55 $\pm0.04$  $\pm0.02$   &  --      &  12.1 $^{+0.2} _{-0.3}$  $\pm0.4$   &   136  &  100 \\ 
Large  &  PL  &  0.836 $\pm0.011$  $\pm0.001$   &  2.36 $\pm0.02$  $\pm0.001$   &  --   &  12.4 $\pm0.2$  $\pm0.4$   &   338  &  149 \\ 
Large &  CPL  &  0.782 $\pm0.014$  $\pm0.005$   &  2.15 $\pm0.04$  $\pm0.02$   &  18.8 $^{+4.2} _{-3.0}$  $^{+2.6} _{-2.1}$   &  11.9 $^{+0.2} _{-0.3}$  $\pm0.4$   &   248  &  148 \\ 
Large  & CPL  &  0.740 $\pm0.007$  $\pm0.002$   &  2; fixed    &  11.4 $\pm0.5$  $\pm0.2$   &  11.6 $^{+0.2} _{-0.3}$  $\pm0.4$   &   280  &  149 \\ 
Large  &  ZA07  &  0.752 $\pm0.008$  $\pm0.002$   &  --   &  1.14 $\pm0.06$  $\pm0.02$   &  11.8 $^{+0.2} _{-0.3}$  $\pm0.4$   &   251  &  149 \\ 
 \hline
 box (a)  &  ZA07  &  0.754 $\pm0.020$  $\pm0.001$   &  --   &  1.90 $^{+0.45} _{-0.32}$  $\pm0.01$   &  1.26 $^{+0.05}_{-0.13}$  $\pm0.02$   &   151  &  136 \\ 
  box (b)  &  ZA07  &  0.775 $\pm0.023$  $\pm0.001$   &  --   &  0.923 $^{+0.136} _{-0.110}$  $^{+0.009} _{-0.008}$   &  1.02 $^{+0.05} _{-0.10}$  $\pm0.03$   &   164  &  120 \\ 
 box (c)  &  ZA07  &  0.825 $^{+0.021} _{-0.020}$  $\pm0.001$   &  --   &  1.03 $^{+0.14} _{-0.11}$  $\pm0.01$   &  1.45 $^{+0.05} _{-0.09}$  $\pm0.04$   &   151  &  150 \\ 
 box (d)  &  ZA07  &  0.761 $^{+0.020} _{-0.019}$  $\pm0.001$   &  --   &  1.26 $^{+0.20} _{-0.16}$  $\pm0.01$   &  1.02 $^{+0.05} _{-0.09}$  $\pm0.02$   &   167  &  141 \\ 
 box (e)  &  ZA07  &  0.663 $\pm0.021$  $\pm0.001$   &  --   &  1.44 $^{+0.31} _{-0.23}$  $\pm0.02$   &  1.14 $^{+0.06} _{-0.12}$  $\pm0.04$   &   121  &  113 \\ 
 box (f)  &  ZA07  &  0.490 $^{+0.022} _{-0.021}$  $\pm0.005$   &  --   &  0.598 $^{+0.093} _{-0.075}$  $^{+0.030} _{-0.025}$   &  1.55 $^{+0.10} _{-0.18}$  $\pm0.16$   &   191  &  131 \\ 
 \hline
\enddata
\tablenotetext{\dagger}{Using only \chandra\ data. Flux is calculated in 0.5--7 keV.}
\tablenotetext{\ddagger}{Using only \nustar\ data. $N_H$ is fixed to the value obtained from \chandra\ spectrum.}
\tablenotetext{\ast}{Flux is calculated by using FPMA. The difference between FPMA and FPMB is smaller than the statistic error. }
\tablecomments{
The errors refer to 90\% confidence level.
The first and second errors of the flux indicate the statistic and the systematic uncertainties, respectively.
}  
\label{tab:fit}
\end{deluxetable*}

\section{Discussion}  \label{sec:4}
\begin{figure}[t!]
\begin{center}
\plotone{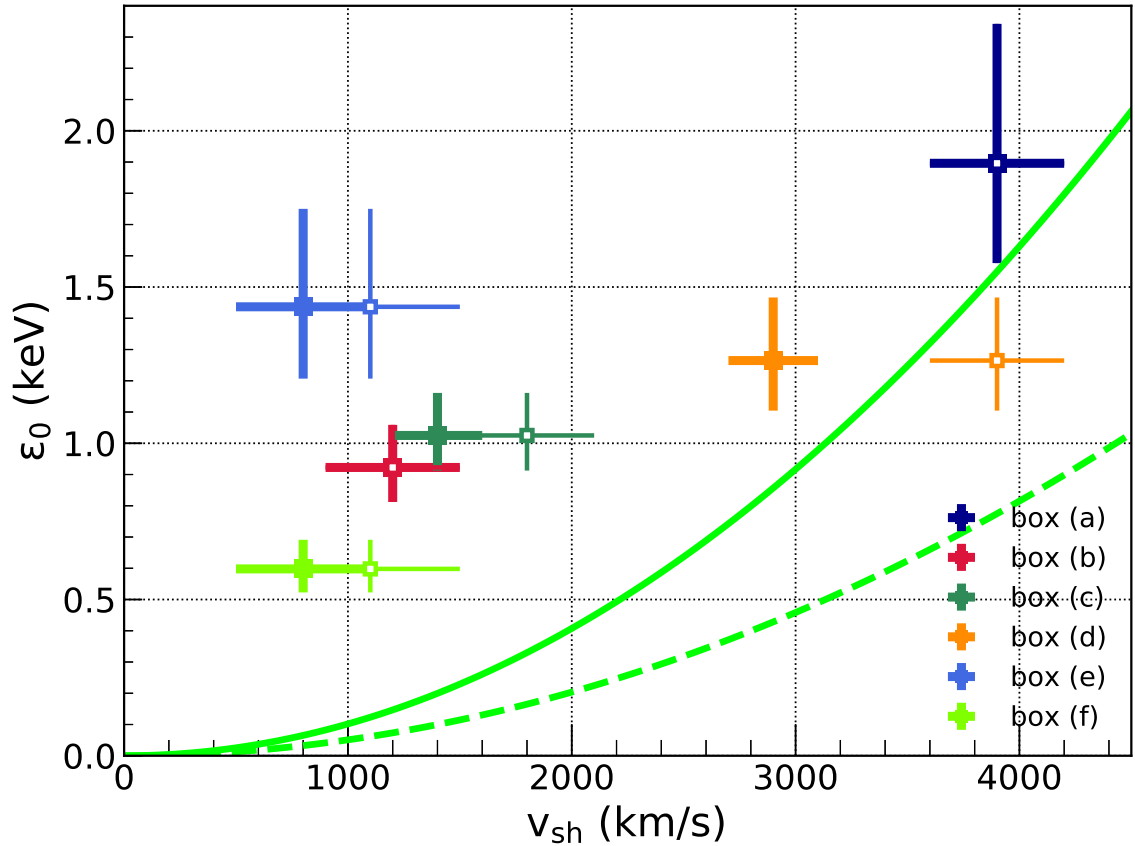}
 \caption{
Relation between shock velocity and cutoff energy parameter.
The error of the cutoff energy parameter shown here is only statistical. Note that the systematic error is much smaller.
The green lines indicate the theoretical relation (\eqref{eq:ZA07_e0}) proposed in \cite{ZA07}. 
Solid and dashed lines are the case of $\eta$=1 (Bohm limit) and $\eta=2$, respectively.
Open markers show the projection-corrected velocities for box (c)--(f) assuming a spherical shell with the radius of 30\arcmin\ i.e. the position of box (a).
} 
\label{fig:cutoff}
\end{center}
\end{figure}

The cutoff energy in the synchrotron spectrum may contain the key information about the mechanism responsible for acceleration of the parent particles.
\cite{ZA07} obtained the following relation between the shock speed and the cutoff energy parameter:
\begin{eqnarray}
\varepsilon_0 = 1.55 \left(\frac{v_{\rm sh}}{3900 ~{\rm km}~{\rm s}^{-1} }\right)^2 \eta^{-1} ~{\rm keV} ,    
\label{eq:ZA07_e0}
\end{eqnarray}
where $\eta\geq1$ is the parameter to indicate the deviation from Bohm limit, so-called gyro factor.
The $\eta=1$ case, known as Bohm limit, is the most efficient acceleration.
The theoretical relation, \eqref{eq:ZA07_e0}, is shown by the green line in \figref{fig:cutoff}. 

We assume the ratio of the upstream magnetic field to the downstream magnetic field, $\kappa = B_{\rm up} / B_{\rm down}$, is $\sqrt{1/11}$.
This indicates an enhancement of random isotropic magnetic field due to the standard shock compression with ratio of $\sigma=4$.
In the case of the equal value of the magnetic field upstream and downstream, $\kappa=1$, Equations (1)--(3) are slightly modified \citep{ZA07}.
Using the models for the $\kappa=1$ case, we also perform the same spectral analysis to the broadband X-ray observations with \chandra\ and \nustar.
The observed $\varepsilon_0$ in $\kappa =1$ case is smaller  than that in $\kappa = \sqrt{1/11}$ case by a factor of 0.6--0.7,
while the theoretically predicted $\varepsilon_0$ is smaller by a factor of 0.6.
This results in the similar trend between the shock velocity and the cutoff energy parameter to the case of $\kappa = \sqrt{1/11}$ illustrated in \figref{fig:cutoff},
which indicates the boxes (a) and (d) are roughly consistent with the theoretical curve, and the rest boxes are not.

As shown in \figref{fig:cutoff}, $v_{\rm sh}-\varepsilon_0$ relation of box (a) and box (d) can be explained by the theoretical relation with $\eta \sim$1.
TeV-scale electrons are accelerated at the maximum rate (Bohm limit) 
in these outermost regions that are likely just behind the forward shock. 
Note that box (d), which seems to be located inside the shell, is possibly indicative of the projected forward shock (box (a)) because the projection-corrected speed is compatible with that of box (a).

On the other hand, the regions with the lower speed (boxes (b), (c), (e), and (f)) do not match the theoretical prediction. 
This suggests that the present framework, which we assume the synchrotron radiation from electrons accelerated at SNR shock through the standard DSA mechanism and limited by the synchrotron loss, is not applicable in these cases.
The inner filament and/or edge at boxes (c), (e), and (f) may represent locally enhanced magnetic fields rather than the acceleration sites. 
Although box (b) exists at the outermost edge, its slow speed and its non-radial direction imply that the shock is decelerated and distorted due to interacting with the molecular clouds \citep{Fukui2003,Fukui2012,Sano2015}. 
Therefore box (b) might not be the acceleration site.

The magnetic field amplifications have been confirmed in the filaments and the small knot-like structures in the previous work of some young SNRs.
Estimated from the width of the filamentary structure of the SNR rim just behind the shock wave,
$B$ is approximately 100 \uG\ in these sub-parsec regions, 0.01--0.4 pc, \citep{Bamba2003,Berezhko2006}.
The magnetic field is expected to be more enhanced, $B\sim$1 mG, in the smaller (0.05 pc) region, derived from year-scale flux variation \citep{Uchiyama2007,Uchiyama2008}.
The box (b), (c), (e), and (f) can be different from these filamentary and small  structures in terms of the location and the size.
The filament-like structure at box (c) and inner edge-like structure between box (e) and (f) are respectively 1.8 pc and 2.6 pc away from the forward shock at box (a).
The size is about 0.4--1 pc.
We might need another scenario for the magnetic field enhancement in these comparably large sub-parsec regions which are isolated from the shock front.

The particle acceleration at reverse shock and/or reflection shock can be feasible for the observed higher cutoff energy parameters and the slower velocities.
Given the position and the speed, the reverse shock does not seem to represent the filament and the inner edge from the hydrodynamical point of view,
although it is not completely excluded \citep{TU16}.
The reflection shock, resulting from the interaction between the SNR shock and the ambient molecular cloud, is reasonable in the case of \rxj\ NW, as discussed in \cite{Okuno2018}.
The evidence of the molecular cloud and the dense clump has been reported \citep{Fukui2003,Fukui2012,Sano2015}.
The shock-cloud interaction causes the deceleration of the shock, which is consistent with the measured slow speeds, and the magnetic field amplification \citep{Inoue2012}.
However it is not enough to consider only the standard acceleration at the reflection shock.
If there exists the reflection shock at the filament or the inner edge, it encounters the ejecta that is freely expanding outward.
In the rest frame of the reflection shock, the upstream velocity is given by
\begin{eqnarray}
u_1 = \frac{R_{\rm ref}}{t_{\rm age}} - v_{\rm obs} ,
\label{eq:u1}
\end{eqnarray}
where $R_{\rm ref}$, $t_{\rm age}$, and $v_{\rm obs}$ are the position of the reflection shock, the SNR age, and the observed apparent velocity, respectively \citep{Sato2018}.
Assuming SN393 is the supernova that created \rxj\ ($t_{\rm age} \approx$1600 years),
$u_1$ is estimated to be $\sim$2700 \kms\ at box (c) and $\sim$2900 \kms\ at box (e).
Using the obtained $u_1$ as the shock speed in \eqref{eq:ZA07_e0}, the observed cutoff energy parameter still appears slightly larger than the theoretical  value.
This also suggests that the standard picture of DSA and the synchrotron cooling is not the case.

The presence of the magnetic turbulence can considerably affect the spectral shape of the synchrotron radiation \citep{ZA10,Bykov2008,Kelner2014}.  The turbulence can be described by different physical quantities, for example, with the power spectrum or with the correlation function. 

In the case of large-scale turbulence, i.e., with the dissipation scale exceeding the photon formation length, the probability distribution of the magnetic field determines the emission spectrum (e.g., Kelner et al. 2014). Derivation of this distribution function from the first principles is a complicated task and is beyond the scope of our paper. 
\cite{ZA10}, however, have obtained an analytical form of the radiation spectrum in the presence of this distribution for the case of turbulence induced by the Weibel instability: 
the cutoff shape of the synchrotron radiation is described by $\exp\left(-\varepsilon^{1/3}\right)$ rather than $\exp\left(-\varepsilon^{1/2}\right)$ (see Appendix in \cite{ZA10}).
Since this seems to be a feasible scenario for SNR shocks, we also utilize their model, 
resulting in that both the obtained cutoff energy parameter and the theoretical prediction are about one order of magnitude smaller than those in the non-turbulent (monochromatic) field.
This eventually produces the almost same trend shown in \figref{fig:cutoff},
where the boxes (a) and (d) are well described with the theoretical curve, and the rest are not.

On the other hand, there is a general relation that describes the impact of turbulence on the diffusion coefficient. We utilize this relation, which links the power spectrum slope and the energy dependence of the diffusion coefficient, to obtain two characteristic energy-dependences of the diffusion coefficient expected in the case of hydrodynamic and MHD turbulence.
The analytical expressions, Equations (1)--(3), are derived in the case of Bohm diffusion,
where the diffusion coefficient is proportional to the energy of particle, i.e. $D(E) \propto E^\alpha$ with $\alpha=1$.
In this case, the electron spectrum has the exponential cutoff form of $\exp(-E^2)$ and the corresponding synchrotron spectrum has $\exp(-\sqrt{\varepsilon})$.
If the diffusion coefficient is deviated from the Bohm diffusion, i.e. $\alpha \neq 1$, the cutoff shape becomes somewhat different.
We can diagnose the diffusion type from the determination of the cutoff shape \citep{ZA07,Yamazaki2014}.
This is beyond the scope of this paper and will be discussed in future publications.

\section{Conclusions} \label{sec:5}
We present the first result of the hard X-ray observations of the NW rim of SNR \rxj\ with \nustar.
The observations are heavily contaminated by the stray lights from nearby X-ray sources and the nonuniform stray-light contribution of CXB and GRXE. 
Using nuskybgd and adding the component of GRXE, we carefully estimate the complex background.
For the first time, the spatially-resolved non-thermal emission up to 20 keV is detected from the remnant.
The morphology taken with \nustar\ is broadly similar to the soft image in the previous work, except for HXC that is observed in the 10--20 keV image of P1 epoch.
We investigate the spectral distribution in the NW shell with the analytical expression of the synchrotron radiation in the vicinity of SNR shock.
The obtained cutoff energy is somewhat variable, from 0.6 to 1.9 keV.
The particle acceleration is required to proceed in the regime close to Bohm limit near the forward shock.
On the other hand, we need another scenario to explain the higher cutoff energy than theoretically predicted in the regions with slow speeds, such as the inner edge and the filamentary structure.
The parsec-scale amplification of the magnetic field and/or the acceleration at the reflection shock might be the case.

\acknowledgments
We thank Kaya Mori and Daniel R. Wik for the helpful advice about \nustar\ analysis.
We also thank Dmitry Khangulyan for the fruitful discussion.
This work was made use of data from the \nustar\ mission, a project led by the California Institute of Technology, managed by the Jet Propulsion Laboratory, and funded by the National Aeronautics and Space Administration. 
This work is supported by the \nustar\ Cycle 1 observation program.
We appreciate the \nustar\ Operations, Software, and Calibration teams for support with the execution and analysis of these observations.
N.T. is supported by the Japan Society for the Promotion of Science (JSPS) KAKENHI Grant Number JP17J06025.
R.K. acknowledges support from the Russian Science Foundation (grant 19-12-00369).
This work was partially supported by JSPS KAKENHI Grant Numbers JP18H03722.
\software{NuSTARDAS (v1.4.1), HEAsoft (v6.19), nuskybgd (https://github.com/NuSTAR/nuskybgd; \cite{Wik2014}), XSPEC (v12.9.0, \cite{Arnaud1996}), CIAO (v4.9, \cite{Fruscione2006}) }




\end{document}